\documentclass{article}

\usepackage{charter,lipsum,fancyvrb,ragged2e}

\usepackage{amsfonts}
\usepackage{amsmath}
\usepackage{pdflscape}
\usepackage{afterpage}

\usepackage{mathrsfs}
\usepackage{multirow}
\usepackage[table]{xcolor}
\usepackage{setspace}
\usepackage{graphicx}
\usepackage[utf8x]{inputenc}
\usepackage{newunicodechar}
\usepackage{hyperref}
\usepackage{color}
\usepackage{caption}
\usepackage[mathscr]{euscript}
\usepackage{amsthm, amssymb, amsfonts}
\usepackage{txfonts}
\usepackage{soul}
\usepackage{stmaryrd}
\usepackage{setspace}
\usepackage{txfonts}
\usepackage[ruled]{algorithm2e}
\usepackage{tabto}
\usepackage{algpseudocode}
\usepackage{pifont}

\usepackage{tcolorbox}
\usepackage{comment}
\usepackage{arydshln}

\usepackage[english]{babel}
\usepackage{multirow}
\usepackage{imakeidx}
\makeindex
\usepackage{array}

\usepackage{subfig}
\usepackage{wasysym}
\usepackage{authblk}
\usepackage{tikz}

\theoremstyle{definition}

\newcolumntype{M}[1]{>{\centering\arraybackslash}m{#1}}
\newcolumntype{N}{@{}m{0pt}@{}}

\DeclareSymbolFont{rsfs}{U}{rsfs}{m}{n}
\DeclareSymbolFontAlphabet{\mathscrsfs}{rsfs}

\RequirePackage{setspace}

\theoremstyle{definition}

\definecolor{antiquewhite}{rgb}{0.98, 0.92, 0.84}
\definecolor{babyblueeyes}{rgb}{0.94, 0.97, 1.0}

\newcounter{exam}
   
   \newcommand{\specialcell}[2][c]{%
  \begin{tabular}[#1]{@{}c@{}}#2\end{tabular}}

\addto{\captionsenglish}{%
}

\providecommand{\keywords}[1]
{
  \small	
  \textbf{\textit{Keywords---}} #1
}

\begin{document}

\title{Database Intrusion Detection Systems (DIDs): Insider Threat Detection via Behavioural-based Anomaly Detection Systems - A Brief Survey of Concepts and Approaches }

\author[$\dagger$]{Muhammad Imran Khan} 
\author[$\ddag$]{Simon N. Foley}
\author[$\dagger$]{Barry O'Sullivan}

\affil[$\dagger$]{Insight Centre for Data Analytics, School of Computer Science and Information Technology, University College Cork, Ireland.}
\affil[$\ddag$]{Department of Information Security and Communication Technology, Norwegian University of Science and Technology, Gj\o{}vik, Norway.}

\date{}

\maketitle

\begin{abstract}
One of the data security and privacy concerns is of insider threats, where legitimate users of the system abuse the access privileges they hold. 
The insider threat to data security means that an insider steals or leaks sensitive personal information. 
Database Intrusion detection systems, specifically behavioural-based database intrusion detection systems, have been shown effective in detecting insider attacks. 
This paper presents background concepts on database intrusion detection systems in the context of detecting insider threats and examines existing approaches in the literature on detecting malicious accesses by an insider to Database Management Systems (DBMS).
\end{abstract}


\section{Introduction}
The recent past has witnessed an exponential increase in the volume of data being stored and accessed through Database Management Systems (DBMS) and this comes along security and privacy concerns.
Organizations need to take extra care in the management and storage of sensitive application data.
Misuse or leakage of such data can lead to an organization suffering from damages in terms of reputation and financial loss. 
The harm caused by data breaches has routinely been reported in the popular press. 

Threats to an organization's data come from external attackers - \textit{outsider attacks} or internal attackers - \textit{insider attacks}.
Traditional security controls such as authentication, role-based access control, data encryption and physical-security can help to control access to this data.
However, there is a persistent concern of insider attacks whereby legitimate users of the system abuse the access privileges they hold.
Therefore, effective security controls to mitigate insider attacks are desirable. 

An Intrusion Detection Systems (IDS), in particular, behavioural-based intrusion detection systems (also referred to as behavioural-based anomaly detection systems)~\cite{10,12,17}, can play a role in detecting insider attacks.
Behavioural-based anomaly detection systems model normative behaviour of the system user and look for deviations in the run-time behaviour of the users.

This paper examines existing literature on detecting malicious accesses by an insider as anomalies, we refer to these anomalies as security-anomalies. 
Section~\ref{21} considers the threats to contemporary organizations, while Section~\ref{22} and ~\ref{23} defines and discusses the impact of an \textit{insider threat}. 
Section~\ref{24} explores the detection methods for insider attacks.
Section~\ref{25} presents a taxonomy of anomaly-based detection methods and reviews the well-known techniques reported in the literature.

\section{Threats to Contemporary Organizations} \label{21}
Threats to an organization can be classified as external (outsider attack) or internal (insider attack).
External threats come from attackers outside of the organization who discover network and/or system vulnerabilities and use this information to penetrate the organization. 
Outside attackers may, for example, utilize social engineering techniques to accomplish a malicious goal, such as stealing confidential information or making some resources unavailable using a Denial-of-Service attack. 
Much research exists on dealing with external threats, and many security defences have been proposed, including host-based access controls, Intrusion Detection Systems (IDSs), and access control mechanisms~\cite{104, 105, 106}.
On the other hand, an insider is a person who belongs to an organization and is authorized to access a range of its data and services.
We are particularly interested in insider attacks as the nature of these attacks make them challenging to detect. 
The next section reviews how the understanding and definition of insider has evolved in the literature and provides the definition used in this work.

\subsection{Defining Insiders} \label{22}
Several definitions can be found in the literature for an insider~\cite{110,111,134}; however, there is no consensus for a single definition~\cite{107}. 
In the 2008 paper, ``Defining the Insider Threat'', Bishop and Gates considered three definitions of insiders~\cite{110}. 
The first definition was from a RAND report~\cite{108} that defines an insider to be \textit{``an already trusted person with access to sensitive information and information systems''}. ~\index{RAND report}
The second definition was also from the same RAND report, which defines an insider to be \textit{``someone with access, privilege, or knowledge of information systems and services''}.
The third definition, originating from~\cite{109}, defines an insider to be \textit{``anyone operating inside the security perimeter''}. 
In the first definition, a person needs to be trusted in order to be called an insider, however, in the second definition, a person having knowledge of the system and services is also considered as an insider while the third definition considers everyone within the security perimeter to be an insider.

The first three definitions are regarded as binary definitions because if the person satisfies one of these definitions, then that person is called an insider, otherwise, the person is not an insider. 
Bishop and Gates~\cite{110} also provided a fourth, non-binary, definition for an insider. 
The non-binary notion of an insider is based upon a measure of the damage that an organization would suffer if entities such as resources, important documents, e-mails, source code, etc. are compromised or leaked. 
Each entity is assigned an impact value that specifies this measure of damage. 
Entities with the same impact value are grouped in protection domain groups. 
These protection domain groups are then paired with groups of users having access to entities in protection domain groups. 
Users having access to protection domain groups with the highest impact-value pose the highest risk of insider threat. 
This model provides a spectrum on which the degree to which an insider poses a threat can be identified.
We believe that such a model is useful in developing more fine-grained security mechanisms by taking into account the threat-level that an insider poses. 
This proposed model for insider threat that Bishop and Gates presented is useful in understanding how threats may be traced and aggregated through a system. 
However, the model does not define what is meant by an insider.

In a 2008 cross-disciplinary workshop on \textit{``Countering Insider Threats''}~\cite{111}, a more specific definition was proposed. 
In this workshop, the insider was defined as \textit{``a person that has been legitimately empowered with the right to access, represent, or decide about one or more assets of the organization's structure''}. 
There is a growing consensus over the definition put forwards in a 2008 cross-disciplinary workshop on \textit{``Countering Insider Threats''}~\cite{111} and therefore we consider this definition for this paper.

A famously reported case of an insider attack was of an employee at an office that issues driving licences. 
The employee exploited access privileges to issue fraudulent licenses~\cite{134}. 
It has been reported that insider attacks can be unintentional as well. 
For instance, the breach resulted from the carelessness of an employee~\cite{94}.

Once we've defined who an insider, we look at the definition of an insider threat is defined.
The definition for insider threat that has consensus on it is put forward by Predd et. al. in~\cite{112} that defines insider threat as \textit{``[...] an insider's action that puts an organization or its resources at risk''}~\cite{112}. 

\subsubsection{Masqueraders and Masquerade Attacks}
\begin{figure}
  \includegraphics[width=\linewidth]{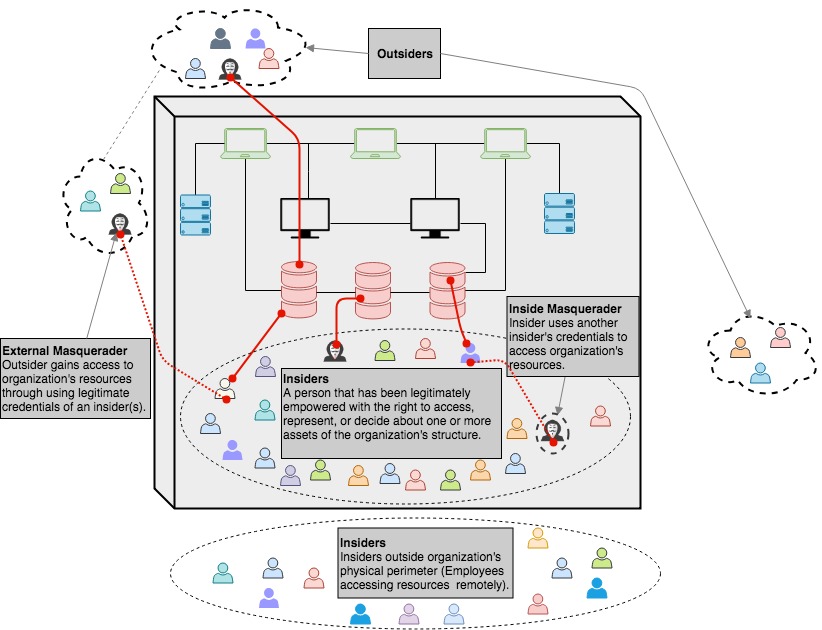}
  \caption{Figure depicting insiders, outsiders, internal and external masqueraders in organizational settings.}
  \label{insider}
\end{figure}

It is worth mentioning that the chosen definition of an insider does not cover the \textit{masqueraders}.
A masquerader is defined as an attacker who has gained (steals) the credentials of an employee (insider) of a contemporary organization and subsequently, uses those credentials to maliciously access organization resources (includes databases) by impersonating as a legitimate employee of that organization. 
A masquerade attack can take one of the two following forms (i) - a masquerader is an insider who gains control of credentials of another employee of the same organization having different privilege level than that being held by the masquerader, (ii) - the masquerader is an outside attacker who somehow gains control of legitimate employee's credentials. 
A distinction, in-terms knowledge base, can be drawn between an \textit{internal masquerader} and \textit{external masquerader}. 
Intuitively, an insider masquerader knows more about the organization as compared to an external masquerader though this is not necessarily always be the case.
Additionally, an internal masquerader can mimic behaviour similar to the behaviour of other employees of the organization, whereas an external masquerader's behaviour is likely to manifest differently because of the lack of knowledge about employee behaviours.
Figure~\ref{insider} a depiction of insiders, outsiders, and  internal and external masqueraders in an organizational setting.

\subsection{The Impact of an Insider Attack} \label{23}
A 2015 report titled \textit{Insider Threat Report} by Vormetric~\cite{32} reported that globally 89\% of the respondent organizations are at risk of insider attack and that among these respondent organizations 34\% of them felt extremely vulnerable to this kind of attack. 
Of the respondent organizations, 56\% plan to increase their spending on tackling the challenge of insider threat. 
The 2015 \textit{Cost of Cyber Crime: Global Report} from the Ponemon Institute~\cite{6} reported that insiders cause the costliest cyber-crimes. 
Information Systems Audit and Control Association (ISACA), reported that, globally, insider threat was among the top three threats for 2016~\cite{113}. 
In the context of healthcare, insiders have reported as the cause of most health-care data breaches~\cite{114}.
For example, a health-care data security report from IBM Managed Security Services reported that insiders were responsible for 68\% of all network attacks targeting health-care data  in 2016~\cite{125}. 
Insider attacks are on the rise, as reported in a 2015 survey report~\cite{7}  that internal factors caused 43\% of data breaches as compared to 2004 where it was reported in~\cite{93} that insiders caused 29\% of the crimes.

Nonetheless, the reporting rate of insider attacks remains very low for a variety of reasons compared to the actual number of instances, including inconsequential impact or lack of evidence~\cite{93}.
Another reason is that organizations are sometimes hesitant to report insider attacks due to loss of reputation and liability;
however, legislations are now in place directing organizations to report data breaches such as the recently passed Privacy Amendment (Notifiable Data Breaches) Act 2016 in Australia~\cite{115} and the EU’s General Data Protection Regulation that came into effect from 2018~\cite{53}. 

\section{Anomaly Detection in Systems} \label{24}
Originally, IDSs were designed to detect network intrusions and can be classified into misuse detection systems and anomaly detection systems~\cite{13}.
Misuse detection systems look for existing misuse patterns and are limited to detect previously known attacks~\cite{13,121, 122}.
In general, misuse detection systems have also been referred to as \textit{signature-based systems} or \textit{knowledge-based intrusion detection systems} in the literature~\cite{129,130,131}.
The majority of commercial intrusion detection systems are misuse detection systems~\cite{188,189,190}. 
Misuse detection systems are circumvented by sophisticated attackers targeting an organization as these systems only detect known attacks~\cite{251}.
A misuse detection system detects attacks by comparing the audit trails with the existing attack signatures.
It provides a guarantee that the known attack is detected, but it cannot detect an unknown attack. 
It is difficult to determine attack signatures for all the variants of a particular attack as different ways exist to exploit vulnerability or weakness. 

In contrast to misuse detection systems, anomaly detection (also known as \textit{behavioural-based}) systems look for a deviation from normative behaviour. 
In principle, anomaly detection systems have the potential to detect zero-day attacks -- attacks for which there is not a known predefined pattern. 
However, in practice, it is a challenge to model normative behaviour accurately. 
Existing work has generally focused on identifying anomalous system operations~\cite{118}, malicious network events~\cite{119} or malicious application system events~\cite{120}.
The anomaly detection systems can be distinguished on the basis of the way in which normative behaviour is modelled, that is,  either the system learns the normative behaviour by automatically mining the past behaviours (learning-based anomaly detection systems) or the normative behaviour (specification-based anomaly detection systems) is specified manually~\cite{132,133}.

It has been reported that the attacker can evade the anomaly detection system by carefully mimicking normative behaviour while exploiting a vulnerability. 
Such attacks are known as \textit{mimicry attacks}~\cite{252,253,254,255,256}.

The effectiveness of IDSs is evident in domains like computer networking, operating systems, and industrial control systems thus making them a favourable choice to be deployed to protect databases against intrusions~\cite{14,80,81,240,241,242,243,244,245,246,247,248,249}.  
However, IDSs deployed to protect systems in the above-mentioned domains are not adequate for databases; therefore, IDSs tailored to databases are desirable.
Such a tailored IDS for DBMS is known as \textit{Database Intrusion Detection System (DIDS)}. 
The literature has shown that intrusion detection systems tailored to databases are effective in the detection of these malicious queries made by an insider~\cite{12,19,18,250}. 
The following sections of this paper review DIDS research and proposes a taxonomy of IDS in the context of DBMS.

An anomaly-based detection system is further classified into learning-based or specification-based.
While remaining within the scope of detecting malicious access to DBMS, the literature lacks any specification-based detection system.
A na\"{i}ve way to design a specification-based detection system would be to list all the legitimate SQL queries.
However, it is impractical to a priori specify every potentially legitimate query. 
The development of a complete specification in the case of DBMS is unattainable, essentially for the inherent flexibility of SQL, that is, a SQL statement can be written in different ways to query the information.
In our opinion, the notion of the specification-based detection system is immaterial in the context of databases intrusion detection because of its impracticality.
In the literature in general, within the context of DIDS research, the anomaly-based detection systems imply that it is a learning-based system.
In literature and commercially, the DIDS solutions are routinely referred to as Data Loss Prevention (DLP) solutions.

\section{A Taxonomy for DBMS Anomaly Detection} \label{25}  
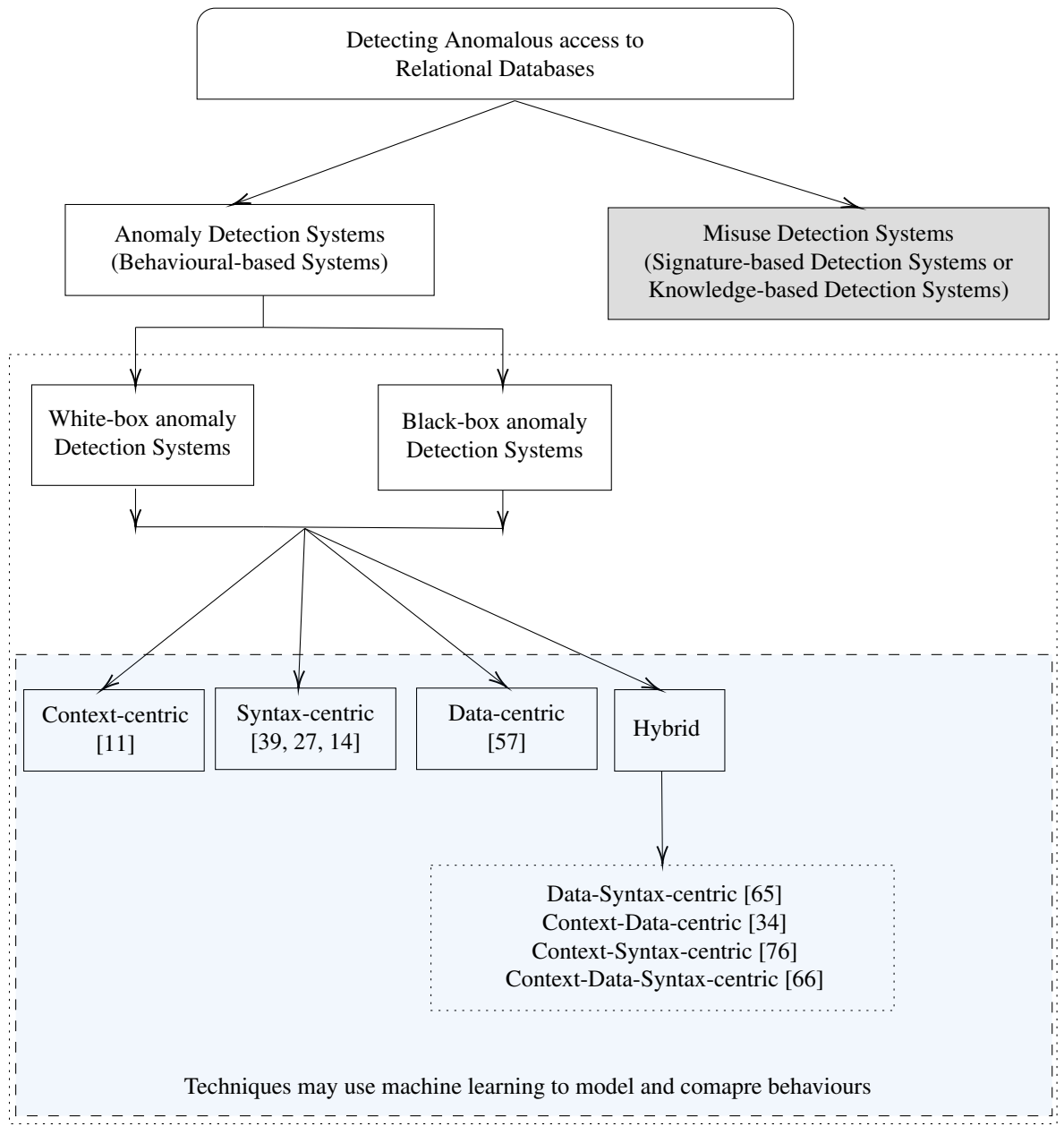
\begin{figure}[h]
\centering
\tikzset{every picture/.style={scale=0.9}} 

\begin{tikzpicture}[x=0.75pt,y=0.75pt,yscale=-1,xscale=1]

\draw   (119,76.4) .. controls (119,70.1) and (124.1,65) .. (130.4,65) -- (481.1,65) .. controls (487.4,65) and (492.5,70.1) .. (492.5,76.4) -- (492.5,122) .. controls (492.5,122) and (492.5,122) .. (492.5,122) -- (119,122) .. controls (119,122) and (119,122) .. (119,122) -- cycle ;
\draw   (36.5,188) -- (267.5,188) -- (267.5,245) -- (36.5,245) -- cycle ;
\draw  [color={rgb, 255:red, 0; green, 0; blue, 0 }  ,draw opacity=1 ][fill={rgb, 255:red, 155; green, 155; blue, 155 }  ,fill opacity=0.34 ] (376.5,190) -- (652.5,190) -- (652.5,260) -- (376.5,260) -- cycle ;
\draw   (10.5,492) -- (123.5,492) -- (123.5,543) -- (10.5,543) -- cycle ;
\draw   (130.5,491) -- (243.5,491) -- (243.5,542) -- (130.5,542) -- cycle ;
\draw   (256.5,491) -- (369.5,491) -- (369.5,542) -- (256.5,542) -- cycle ;
\draw   (380.5,492) -- (449.5,492) -- (449.5,543) -- (380.5,543) -- cycle ;
\draw    (410,543) -- (410.48,599) ;
\draw [shift={(410.5,601)}, rotate = 269.51] [color={rgb, 255:red, 0; green, 0; blue, 0 }  ][line width=0.75]    (10.93,-3.29) .. controls (6.95,-1.4) and (3.31,-0.3) .. (0,0) .. controls (3.31,0.3) and (6.95,1.4) .. (10.93,3.29)   ;
\draw  [dash pattern={on 0.84pt off 2.51pt}] (265.5,602) -- (555.5,602) -- (555.5,693) -- (265.5,693) -- cycle ;
\draw    (318,123) -- (143.38,187.31) ;
\draw [shift={(141.5,188)}, rotate = 339.78] [color={rgb, 255:red, 0; green, 0; blue, 0 }  ][line width=0.75]    (10.93,-3.29) .. controls (6.95,-1.4) and (3.31,-0.3) .. (0,0) .. controls (3.31,0.3) and (6.95,1.4) .. (10.93,3.29)   ;
\draw    (318,123) -- (531.59,189.41) ;
\draw [shift={(533.5,190)}, rotate = 197.27] [color={rgb, 255:red, 0; green, 0; blue, 0 }  ][line width=0.75]    (10.93,-3.29) .. controls (6.95,-1.4) and (3.31,-0.3) .. (0,0) .. controls (3.31,0.3) and (6.95,1.4) .. (10.93,3.29)   ;
\draw    (186.5,391) -- (61.07,490.76) ;
\draw [shift={(59.5,492)}, rotate = 321.51] [color={rgb, 255:red, 0; green, 0; blue, 0 }  ][line width=0.75]    (10.93,-3.29) .. controls (6.95,-1.4) and (3.31,-0.3) .. (0,0) .. controls (3.31,0.3) and (6.95,1.4) .. (10.93,3.29)   ;
\draw    (186.5,391) -- (182.58,489) ;
\draw [shift={(182.5,491)}, rotate = 272.29] [color={rgb, 255:red, 0; green, 0; blue, 0 }  ][line width=0.75]    (10.93,-3.29) .. controls (6.95,-1.4) and (3.31,-0.3) .. (0,0) .. controls (3.31,0.3) and (6.95,1.4) .. (10.93,3.29)   ;
\draw    (186.5,391) -- (310.93,488.76) ;
\draw [shift={(312.5,490)}, rotate = 218.16] [color={rgb, 255:red, 0; green, 0; blue, 0 }  ][line width=0.75]    (10.93,-3.29) .. controls (6.95,-1.4) and (3.31,-0.3) .. (0,0) .. controls (3.31,0.3) and (6.95,1.4) .. (10.93,3.29)   ;
\draw    (186.5,391) -- (406.68,491.17) ;
\draw [shift={(408.5,492)}, rotate = 204.46] [color={rgb, 255:red, 0; green, 0; blue, 0 }  ][line width=0.75]    (10.93,-3.29) .. controls (6.95,-1.4) and (3.31,-0.3) .. (0,0) .. controls (3.31,0.3) and (6.95,1.4) .. (10.93,3.29)   ;
\draw   (15.5,301) -- (154.5,301) -- (154.5,364) -- (15.5,364) -- cycle ;
\draw   (232.5,301) -- (378.5,301) -- (378.5,367) -- (232.5,367) -- cycle ;
\draw    (160.5,245) -- (160.5,265) ;
\draw    (80.5,265) -- (160.5,265) ;
\draw    (160.5,265) -- (310.5,266) ;
\draw    (80.5,265) -- (80.5,300) ;
\draw [shift={(80.5,302)}, rotate = 270] [color={rgb, 255:red, 0; green, 0; blue, 0 }  ][line width=0.75]    (10.93,-3.29) .. controls (6.95,-1.4) and (3.31,-0.3) .. (0,0) .. controls (3.31,0.3) and (6.95,1.4) .. (10.93,3.29)   ;
\draw    (310.5,266) -- (310.5,300) ;
\draw [shift={(310.5,302)}, rotate = 270] [color={rgb, 255:red, 0; green, 0; blue, 0 }  ][line width=0.75]    (10.93,-3.29) .. controls (6.95,-1.4) and (3.31,-0.3) .. (0,0) .. controls (3.31,0.3) and (6.95,1.4) .. (10.93,3.29)   ;
\draw    (80.5,390) -- (160.5,390) ;
\draw    (160.5,390) -- (310.5,391) ;
\draw    (310.5,367) -- (310.5,389) ;
\draw [shift={(310.5,391)}, rotate = 270] [color={rgb, 255:red, 0; green, 0; blue, 0 }  ][line width=0.75]    (10.93,-3.29) .. controls (6.95,-1.4) and (3.31,-0.3) .. (0,0) .. controls (3.31,0.3) and (6.95,1.4) .. (10.93,3.29)   ;
\draw    (80.5,366) -- (80.5,388) ;
\draw [shift={(80.5,390)}, rotate = 270] [color={rgb, 255:red, 0; green, 0; blue, 0 }  ][line width=0.75]    (10.93,-3.29) .. controls (6.95,-1.4) and (3.31,-0.3) .. (0,0) .. controls (3.31,0.3) and (6.95,1.4) .. (10.93,3.29)   ;
\draw  [fill={rgb, 255:red, 74; green, 144; blue, 226 }  ,fill opacity=0.07 ][dash pattern={on 4.5pt off 4.5pt}] (5.5,470) -- (654.5,470) -- (654.5,759) -- (5.5,759) -- cycle ;
\draw  [dash pattern={on 0.84pt off 2.51pt}] (1.5,282) -- (657.5,282) -- (657.5,764) -- (1.5,764) -- cycle ;

\draw (305.75,93.5) node   [align=center] {Detecting Anomalous access to  \\
Relational Databases 
};
\draw (152,216.5) node   [align=center] {Anomaly Detection Systems \\ (Behavioural-based Systems)};
\draw (514.5,225) node   [align=center] { 
Misuse Detection Systems \\ (Signature-based Detection Systems or \\
Knowledge-based Detection Systems)};
\draw (68,517.5) node   [align=center] {
Context-centric \\ \cite{86}
};
\draw (187,516.5) node   [align=center] {
Syntax-centric \\ \cite{12, 126, 250}
};
\draw (313,516.5) node   [align=center] {
Data-centric \\ \cite{19}
};
\draw (415,517.5) node   [align=center] {
Hybrid 
};
\draw (412,648) node   [align=center] { 
Data-Syntax-centric  \cite{18} \\ Context-Data-centric  \cite{102} \\ Context-Syntax-centric  \cite{48} \\ Context-Data-Syntax-centric  \cite{127}
};
\draw (85,332.5) node   [align=center] { 
White-box anomaly \\ Detection Systems  
};
\draw (305.5,334) node   [align=center] { 
Black-box anomaly \\ Detection Systems 
};
\draw (328,741.5) node   [align=center] { 
Techniques may use machine learning to model and comapre behaviours
};

\end{tikzpicture}

\caption{Taxonomy of anomalous DBMS-access detection systems.
} \index{Taxonomy of anomaly-based DIDS}
\label{Taxonomy}
\end{figure}

This section introduces a taxonomy of methods that detect anomalous access to a DBMS.
Anomaly-based DIDS research has remained a centre of focus of the research community, while less attention is being paid on misuse (or signature)-based DIDSs.
The proposed taxonomy is shown in Figure~\ref{Taxonomy} that broadly categorizes IDSs to detect anomalous access in a DBMS.
An aspect to keep in consideration while performing classification of anomaly-based DIDS is the set of features used to model normative behaviour of a user, for example, time of access, attributes in projection clause, relations/tables queried etc.
Section~\ref{FbD} and its following sections discuss these classifications. 
To address these classifications, we first present the prevalent architecture of anomaly-based database intrusion detection systems in the following section.

\subsection{Prevalent Architecture of Anomaly-based Database Intrusion Detection Systems}

\begin{figure}
\centering
\includegraphics[width=\linewidth]{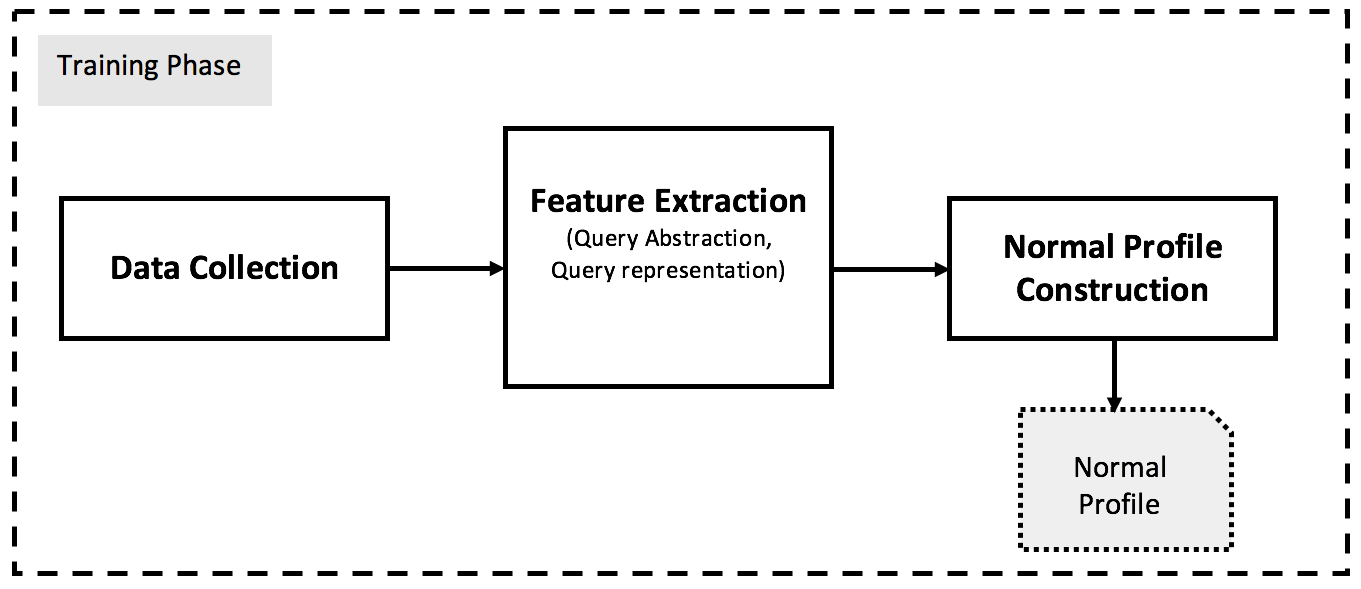}
\caption{Training phase of an anomaly detection system.}
\label{adar1}
\end{figure}

\begin{figure}
\centering
\includegraphics[width=\linewidth]{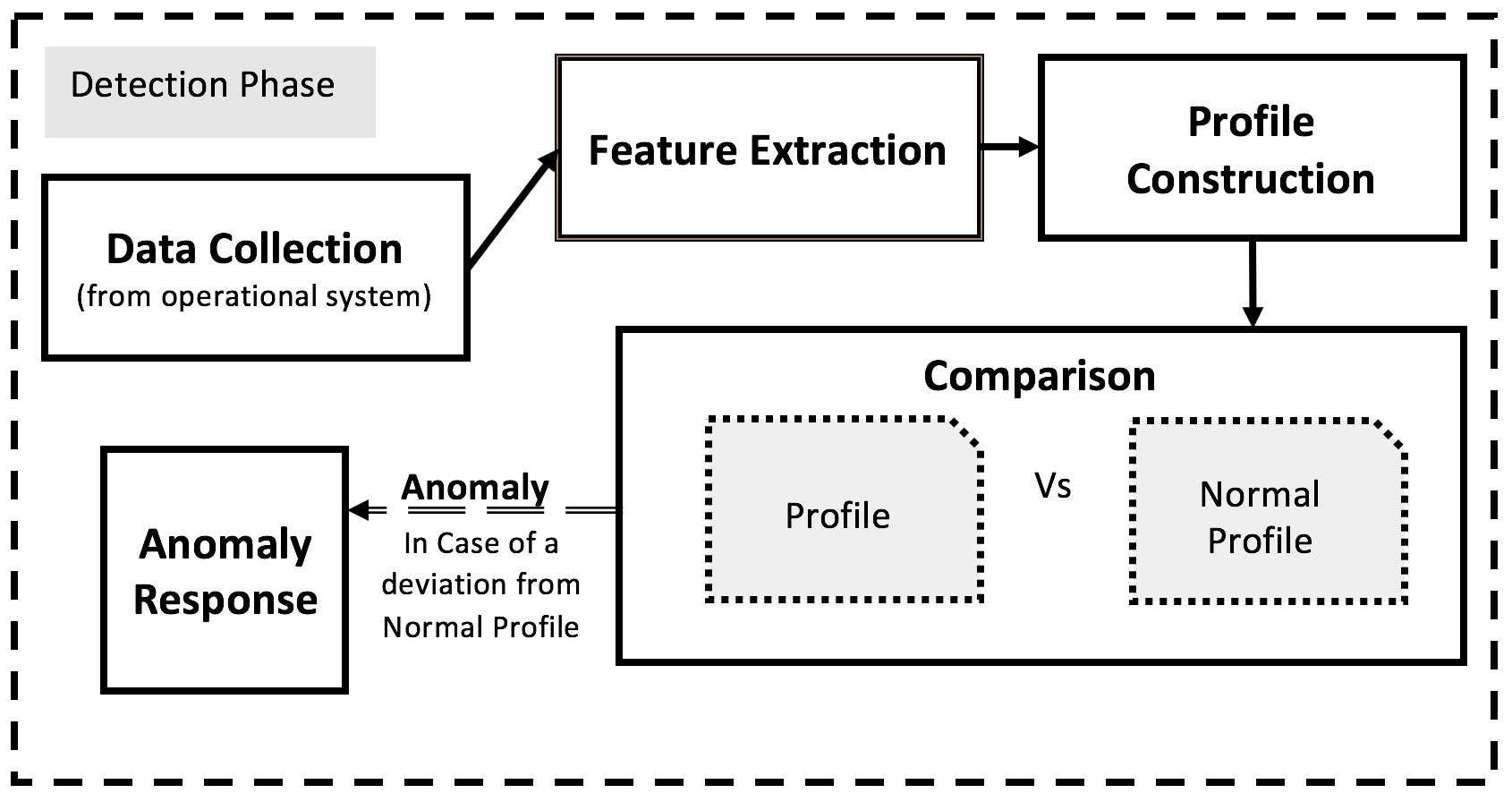}
\caption{Detection phase of an anomaly detection system.}
\label{adar2}
\end{figure}

Figures~\ref{adar1} and \ref{adar2} depict the prevalent architecture of an anomaly-based database intrusion detection system.
The architecture involves two phases: a training phase and a detection phase. 
A profile of normative behaviour is constructed during the training phase and is compared against the run-time profile that is constructed in the detection phase. 
The future deviations of the run-time profile from normative profiles are labelled as anomalies and necessitate attention.
The architecture consists of two fundamental components: feature extraction (abstraction) and profile constructor. 
In feature extraction, the features required to construct the profile are extracted. 
The profile construction component is the technique to construct the profile using selected features. 

Profiling ``normative behaviour'' is non-trivial as many features about queries can be taken into consideration while constructing profiles, and identifying significant determinant features is a challenge.
The classification based on features and the techniques to construct profiles is discussed in the following next sections.

\subsection{Feature Classification} \label{FbD} 
Anomaly-based intrusion detection techniques for detecting malicious access to databases can be further distinguished based on what features they extract from a DBMS  audit log of SQL queries to model behaviours.
The modelled behaviour is represented as a behavioural profile~\cite{10,12}.
These features can be syntax-centric, context-centric, and data-centric, which is sometimes is referred to as result-centric in the literature~\cite{12,18,19}. 

Techniques using syntax-centric features construct behavioural profiles by using syntax features of the SQL query included but are not limited to the attributes in a projection clause, the relations queried, the attributes in selection clause, and/or the type of SQL command~\cite{12}. 
Techniques using data-centric or result-centric features construct behavioural profiles using data returned in response to an SQL query or any other statistical measurement on returned data, for instance, the minimum and maximum value in case of numeric data. 
For example, one could use the amount of information (the percentage of data returned) returned in response to a query or the returned values of attributes to model behaviour or a user~\cite{19}.
The context-centric techniques construct behavioural profiles using contextual features. 
Contextual features  are associated with the context of the query, for example, the time at which the query was made, the user ID of the person making the query, or the number of queries made in a specified time period, etc~\cite{86}. 
A combination of the context, syntax and data-centric features, can be used while modelling normative behaviours.
One such anomaly detection technique that uses syntax and data-centric features is proposed in~\cite{18}. 

An anomaly-based detection system building behavioural profiles that are not understandable by humans in a meaningful way is known as a \textit{Black-Box} anomaly detection system.
On the other hand, an anomaly-based detection system building behavioural profiles that are understandable by humans are known as \textit{White-Box} anomaly detection system.
Understandability implies that the actual root cause of the anomaly can be identified by the administrators (security officer, database administrator, etc.) when they inspect the anomaly. 
Intuitively, white-box approaches have the potential to help explain anomalies.

\subsection{SQL Query Abstraction} \label{SQL:12}
Syntax-centric implies using only factors associated with the syntax of an SQL statement.  
A question that arises is how much of the SQL statement can be considered while constructing a behavioural model, that is, should one consider the entire statement or some parts of the statement -- the challenge of selecting appropriate SQL query abstraction.
Abstraction is a tuple representation of an SQL statement and consists of query features like relation name, attribute names, the amount of returned data or any statistics on the returned data. 
A number of techniques~\cite{12,17,18,40,41,42,43} have been proposed that transform the syntax of an SQL statement into a more abstract fingerprint that can be used for comparing queries. 
SQL query abstractions are also referred to as SQL query fingerprints~\cite{40,41}, SQL query signatures~\cite{12} or SQL query skeletons in the literature~\cite{17}.
The query abstractions used by existing anomaly-based detection approaches are also described in this paper. 

Other than DIDS research, the use of abstraction in practice has also been studied by audit log summarization research as typically audit logs encompass a large number of queries and methods to summarize logs in a meaningful way are sought~\cite{209}.
For instance, it has been reported in a recent study that within the time period of $19$ hours, approximately $17$ million SQL queries were made in a major US bank~\cite{17}.

\subsection{Syntax-centric Features-based Techniques} \label{26}
Several anomaly detection approaches to detect malicious accesses to DBMS use syntax-centric features of the SQL queries to construct behavioural profiles~\cite{12, 126, 250}.
For instance, the approach \textit{DetAnom}~\cite{12} detects malicious DBMS-accesses by application programs. 
SQL queries executed by an application program are represented in the form of SQL query abstractions to generate a normative profile of an application.
In~\cite{12}, a SQL query abstraction consists of the following elements, $(c, t, r, q, n)$, where $c$ is SQL command type, for instance, \texttt{SELECT}. 
Attribute $t$ is the list of attribute identifiers projected in the query and are relative to the relation to which they belong. 
Attribute $r$ is the list of relation identifiers. 
Attribute $q$ is the list of attribute identifiers in the \texttt{WHERE} clause and $n$ is the number of predicates in the \texttt{WHERE} clause. 
For example, consider the following SQL query: \\
\definecolor{darkbrown}{rgb}{0.44, 0.0, 1.0}
\definecolor{deepcarrotorange}{rgb}{0.91, 0.41, 0.17}
\definecolor{coolblack}{rgb}{0.76, 0.33, 0.76}
\begin{flushleft}
\texttt{SELECT \textcolor{black}{bankdb}.\textcolor{black}{acc\_number}, \textcolor{black}{bankdbacc}.\textcolor{black}{current\_amount}}, 
\\  FROM bankdbacc, bankdb, 
\\  WHERE \textcolor{black}{bankdb}.\textcolor{black}{acc\_number}=7594 
\\ AND \textcolor{black}{bankdb}.\textcolor{black}{acc\_number} = \textcolor{black}{bankdbacc}.\textcolor{black}{account};
\end{flushleft}
The corresponding identifiers for relations and the attributes are shown in Table~\ref{ex:ta}.
This query is then represented as follows: \\
  \{$\langle$1$\rangle$    $\langle$\textcolor{black}{100}:\textcolor{black}{10}, \textcolor{black}{200}:\textcolor{black}{20}$\rangle$     $\langle$100, 200$\rangle$    $\langle$\textcolor{black}{100}:\textcolor{black}{20}, \textcolor{black}{200}:\textcolor{black}{30}$\rangle$     $\langle$2$\rangle$ \}.
 
The generation of a query abstraction is followed by the exploration of all execution paths of the application program accomplished by Concolic execution (Concolic testing) which is  a technique for program analysis~\cite{96,97,98}.
Once all the paths are explored, the branching condition is then paired with the SQL query that falls under that branching condition that was discovered in the process of concolic testing. 
This \textit{(branch condition, SQL query)} pair is referred to as a query record. 

In the detection phase of DetAnom, a query is intercepted, and its branch condition is matched with the branch condition in the corresponding a query record. 
If the branch condition is satisfied, then matching of query abstraction is carried out. 
In case of a mismatch, the query is said to be malicious.
DetAnom works only for application programs where the SQL queries are embedded in program code, and it is not extendible to capture users behaviours manifested in a variety of SQL queries made to the database.   

\begin{table}[]
\center
\begin{tabular}{|l|c|}
\hline
\textbf{Command Type}   & \multicolumn{1}{l|}{\textbf{Representation}}       \\ \hline
\texttt{SELECT}         & 1                                         \\ \hline
\texttt{UPDATE}         & 2                                         \\ \hline
\texttt{INSERT}        & 3                                         \\ \hline
\texttt{DELETE}         & 4                                         \\ \hline \hline \hline
\textbf{Relation Name}  & \multicolumn{1}{l|}{\textbf{Relation Identifier}}  \\ \hline 
\texttt{bankdb}        & 100                                       \\ \hline
\texttt{bankdbacc}     & 200                                       \\ \hline \hline \hline
\textbf{Attribute Name} & \multicolumn{1}{l|}{\textbf{Attribute Identifier}} \\ \hline 
\texttt{acc\_number}     & 10                                        \\ \hline
\texttt{current\_amount} & 20                                        \\ \hline
\texttt{account}        & 30                                        \\ \hline
\end{tabular}
\caption{Identifiers example for the query abstraction approach proposed in~\cite{12}.}
\label{ex:ta}
\end{table}

Another early approach based on syntax-centric features is presented in~\cite{250}.
The approach uses three query abstractions each having different level of granularities, coarse triplet, medium-grain triplet, fine triplet. 
A Na\"{i}ve Bayes classifier was used to predict a role for the SQL query.
If the predicted role is not the same as from the one SQL query was originated then an alarm is raised.
This approach has several limitations including, considering each query in isolation and it does not consider the sequence of queries, 
Second, the approach constructed profiles of roles while ignoring the case where one user can belong to multiple roles.

The proposed approach in~\cite{10} demonstrates that one can build behavioural-based anomaly detection systems by considering the sequence of queries (query correlations) to model insider's querying behaviour to detect malicious accesses manifested in sequences of queries rather than a query in isolation, to DBMS.
The approach~\cite{10} models querying behaviour of an insider using n-grams that capture short-terms SQL query correlations. 
The model used abstractions of SQL query audit logs to construct insider profiles, a normative profile using safe logs and a run-time profile using run-time audit logs. 
The run-time profile is compared with normative profile and deviations are an indication of anomalies.

This model proposed in~\cite{288} introduced the notion that behaviours that are rare (infrequent) represent potentially malicious access by an insider, and frequent behaviours are possibly safe behaviours. 
The domain of item-set mining was explored. Item-set mining algorithms including PrePost+, Apriori-Inverse, and Apriori-Rare were adopted to mine frequent and rare query-sets to model querying behaviours (in terms of SQL query abstractions). 
Results point towards the potential effectiveness of modelling insiders malicious querying behaviour as rare behaviour that also enables detection of insider's malicious databases accesses as anomalies.

Syntax-centric approaches, in general, are useful in detection masquerader attacks as well as SQL injection attacks both these attacks lead to structural changes in SQL statement.

\subsection{Data (Result)-centric Features-based Techniques} \label{27}
Little research has been reported on using data-centric features as the basis for anomaly detection in the context of DIDS.
Data-centric features include the amount of data returned in response to a query or returned values of attributes or any other statics performed on the returned set of attribute values. 

In~\cite{19}, it is argued that syntax-centric features of a query alone are a poor discriminator of intent. 
Syntactically different queries can potentially give the same result while syntactically similar queries can potentially yield different results. 
Therefore, a user can craft a legitimate SQL query to retrieve results from the database which the user is authorized to retrieve, yet a purely syntax-centric anomaly-based detection system might label this as anomalous behaviour.

In the approach proposed in~\cite{19}, user profiles are clusters that are specified in terms of an \textit{S-Vector} that provides a statistical summary of the results (tuples/rows).
In the detection phase, clustering algorithms were adopted, that are, as supervised learning methods, Euclidean k-means clustering, Support Vector Machines (SVM), Decision Tree Classifier, and Na\"{i}ve Bayes, and as unsupervised methods, Cluster-Based Outlier Detection (based on Euclidean distance clustering) and Attrib-Deviation, using L$_{\infty}$-norm.
If a query belongs to the cluster, then it is considered normal else it is regarded as anomalous.
The presented approach is suitable for the detection of a query in isolation and does not take a sequence of queries into account.

The approach in~\cite{64} looked at the modelling of behaviour from a DBMS's perspective. 
A record/DBMS oriented approach (also referred to as a semantic approach) is presented in~\cite{64} that considers frequency-based correlations to detect insiders malicious accesses as anomalies. 
The construction of the profiles utilizes control charts from statistical process control as a way to detect anomalies. 
Two scenarios were considered, in the first scenario, the training data for modelling normative behaviour contains outlier. 
In the second scenario, the training data for modelling normative behaviour is free from outliers. 
The experiments demonstrated the effectiveness of the approach in the detection of frequent observation attacks by insiders as anomalies. 
It was discovered that the semantic approach not only identified unseen behaviours but also identified behaviours that should have been present in the current behaviours as anomalies, which we refer to as oversight-anomalies. 
Oversight-anomalies are the anomalies introduced due to human negligence or human errors, for example, an instance where the doctor or the nurse (caregiver) missed a daily check-up
of a patient. 
To the best of our knowledge, this is the first time the oversight-anomalies are considered in the DBMS setting. 
It was also demonstrated that the proposed model for the construction of record-oriented profiles could be transformed into a model for the construction of role-oriented profiles.

Data-centric approaches are capable of detecting sophisticated attacks as well.
For instance, data harvesting attacks involving retrieval of a large amount of data, therefore, exceeding what is retrieved by a legitimate user.

\subsection{Context-centric Features-based Techniques} \label{28}
Few purely context-centric approaches are reported in the literature. 
One such context centric approach is presented in~\cite{86} in which contextual features are considered in modelling user behaviours.
The approach took the deployment of anomaly-based IDS in the medical sector as its use-case and studied the Break-The-Glass (BTG) procedure which is a procedure that breaks the traditional access control mechanism and enable access of patients data in case of emergency to employees of different departments. 
In this approach in~\cite{86}, users who supposed to behave similarly are divided into groups and profiles are constructed for groups. 
The feature space comprised  of contextual features like access type, time, division, date.
Profiles were represented as the sequence of histograms and were constructed using the concept of \textit{Bins}. 
Bins represent the frequency of features. 
In the detection phase, the distance between the histogram of a user and the existing profile is measured, and a larger distance is an indication as an anomaly. 
The approach represented user and group profile in terms of a sequence of histograms of features that can easily be interpreted by a concerned individual like a security officer. 
Therefore, the approach in~\cite{86} can be classified as a white-box anomaly detection approach.

Context-centric approaches typically increase the detection effectiveness of an ID approach when combined with syntax or data-centric approach. 

\subsection{Hybrid Techniques} \label{29}
An example of an approach using data and syntax-centric features to construct behavioural profiles is presented in~\cite{18}. 
Machine learning techniques, in particular, Na\"{i}ve Bayes classifiers and multi-labelling classifiers, were also deployed in the profile generation process.~\index{Na\"{i}ve Bayes Classification}
User profiles are built in the training phase from logs containing user activities.

The approach transforms an SQL query into an SQL query abstraction called a quadruplet. 
A quadruplet $QT(c, P_R, P_A, S_R)$ is composed of data-centric and syntax-centric features including the command type $c$; the list of relations accessed by the query $P_R$; 
the list of attributes accessed by query relative to the relation $P_A$; 
and the amount of selected information from the relation $S_R$. 
This hybrid approach is demonstrated in two settings that is role-based anomaly detection and unsupervised anomaly detection. 
In the detection phase the role of a querier was predicted using a Na\"{i}ve Bayes classifier. 
Multi-labelling classification was used in case of an overlap of roles that results in more than one role. 
If the predicted role is different from the actual role then the query is labelled as anomalous.
In the unsupervised settings, the COBWEB~\cite{100,101} clustering algorithm was selected.
The query is treated as anomalous if a query made by a user falls into a cluster that does not contain any query made by this user.
This approach is promising for the detection of a single malicious query in isolation; however, it ignores sequences of queries while modelling behaviour.

The approach presented in~\cite{102} uses context-centric and data-centric features. 
Normative profiles are constructed by discovering association rules between context-centric features and data-centric features using frequent item-set mining~\cite{103}. 
The basic idea of the approach is to tie the results retrieved by the SQL query with the context in which they were retrieved.  
For instance, a transaction made in London in the morning typically retrieves records for employees of the human resource department. 
Therefore, human resource department employees records are tied with the context that they are normally retrieved in the morning from London.  
In the detection phase, for any incoming query, context-centric features were extracted, and rules conforming to these features are matched, and then the result of the query is matched with the results associated with the retrieved rules.
A drawback of this approach is that large databases result in large profiles. 
Additionally, this approach is too restrictive and less likely to be scalable.
Another drawback of this approach is in general the drawback of context-centric approaches that is the context can be easily mimicked. 

The approach in~\cite{48}, also employed context and syntax-centric features to model behaviours and forms some assumptions for instance that every department in an organization has a unique IP space, employees work in shifts (there are three shifts in a day).
The features collected for modelling include Employee ID, Role ID, time, IP address, Access Type (direct or through an application). 
The approach also records the SQL query associated with the contextual features. 
The profile consists of the probability of each feature's occurrence observed for every user.
In the detection phase, a new transaction is compared against the constructed profiles to check the closest issuer (user) of that transaction.
The transaction is labelled as an anomaly in case the issuer of the transaction is different from the one computed.
This approach also considered Role Hierarchy, meaning, if role \"r$_1$ is above Role {\"r}$_2$ in the hierarchy, then the access privileges of Role \"r$_2$ is a subset of Role \"r$_1$. 
For instance, if a query made by \"r$_1$ is labelled as malicious, but the same query is legitimate for \"r$_2$ then this is not considered as a malicious query.
The focus of this approach is to augment Role-Based Access Control (RBAC) to ensure that the query is made only by the authorized users.
Similar to approaches discussed above, this approach also detects only single malicious DBMS transaction where later in the paper it is argued that a single query may be legitimate, however, a group of them made together might result in malicious or illegitimate action.

\begin{figure*}
\centering
\includegraphics[width=\linewidth]{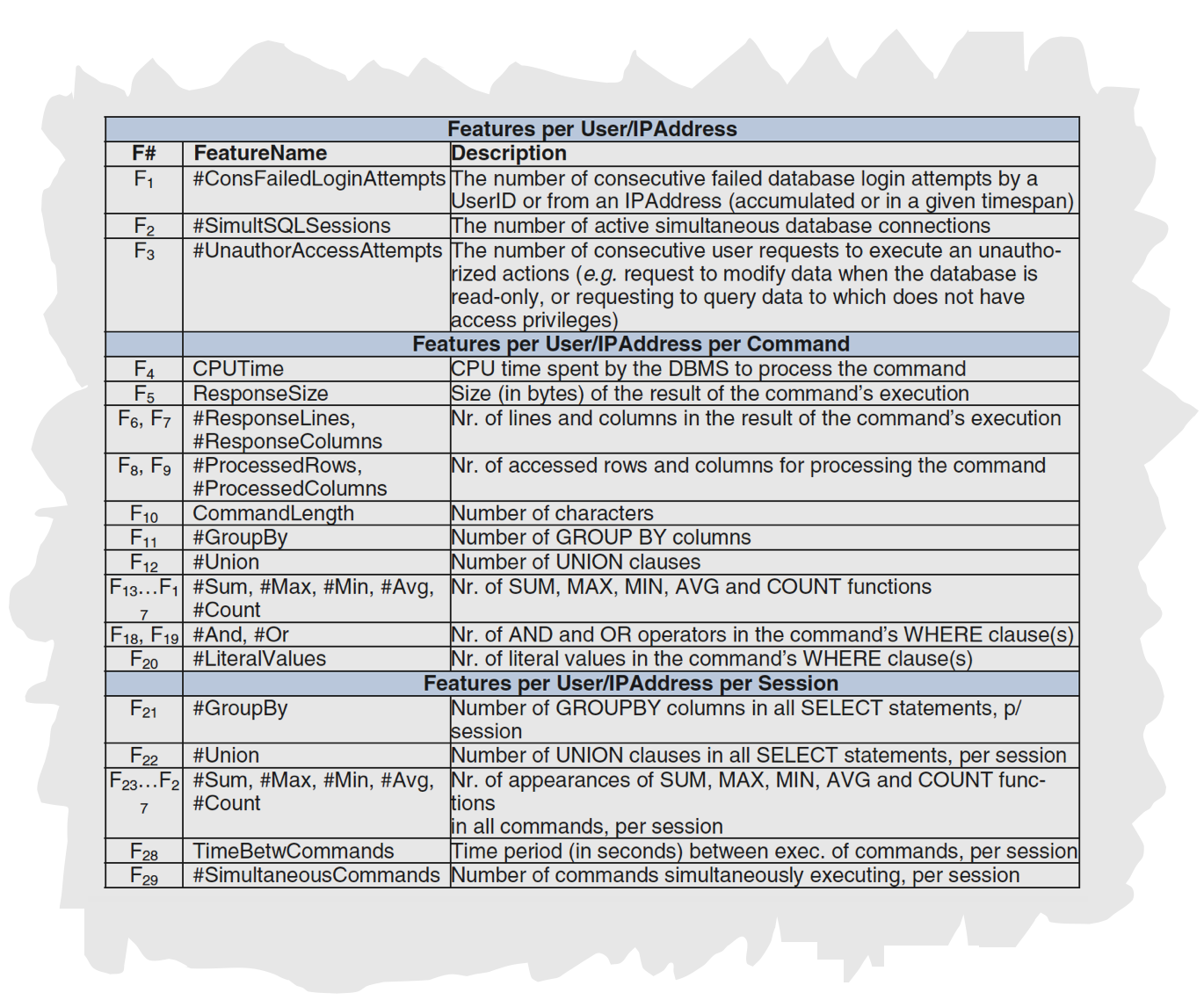}
\caption{List of features considered. Figure cropped from \textit{Securing Data Warehouses from Web-Based Intrusions} by Santos et al.~\cite{127}}
\label{T1}
\end{figure*}

In~\cite{127} contextual, data and syntax-centric features are considered in modelling.
The database intrusion detection approach in \cite{127} is tailored for Data Warehouses in which applications are enabled to access Data Warehouses via the web. 
The profiles are constructed by considering various features, as shown in Figure~\ref{T1} and are represented in terms of the probabilistic distribution of each feature for each user and for the entire population.
In the detection phase for this approach, testing is done to match features distribution with the distributions obtained in the training phase using statistical hypothesis tests like Chi-square, Shapiro-Wilk, and Kolmogorov-Smirnov tests~\cite{127}. 
In case of a non-conformity, the activity involving that feature is labelled as an anomaly.
The approach is focused on web-based malicious access to Data Warehouses though insiders remain unaddressed.
Table~\ref{T2} shows a consolidated presentation of discussed and well-known approaches proposed in the literature.

\afterpage{%
    \clearpage
    \begin{landscape}
        \centering %
\begin{table*} \arrayrulecolor{white}
\begin{tabular}{ccccccc}\hline
        \color{white}
      \cellcolor{black}
      Approach       & \cellcolor{black} \color{white} Features                     & \cellcolor{black} \color{white}  \specialcell{White-box / \\ Black-box} &  \cellcolor{black} \color{white}\specialcell{Single Query \\ detection}  &  \cellcolor{black} \color{white} \specialcell{Sequence\\ of Queries \\ detection} &  \cellcolor{black} \color{white} \specialcell{Profiles} & \cellcolor{black} \color{white} \specialcell{Detection \\ Style} \\ \hline \hline \rowcolor{babyblueeyes}
Hussain et al.~\cite{12}  &       $ \varoast $            & \ding{110}            & \textcolor{green}{\ding{52}} & \textcolor{red}{\ding{54}} &  \specialcell{Tuples of \\SQL-branch condition} &    \specialcell{Tuple matching}              \\ \hline \rowcolor{babyblueeyes}
Khan et al.~\cite{64}  & $\varodot$ & $\Box$            & \textcolor{green}{\ding{52}} & \textcolor{green}{\ding{52}}  &  Record Access Frequency   &    \specialcell{Control Charts} \\ \hline    \rowcolor{babyblueeyes}
Mathew et al.~\cite{19}   & $\varodot $                    & \ding{110}            & \textcolor{green}{\ding{52}} & \textcolor{red}{\ding{54}} &  \specialcell{Clusters / Classes \\ of S-Vector} &      \specialcell{\textbf{Supervised:} \\ Euclidean \textit{k}-mean\\ SVM, Decision Tree Classifier \\ Na\"{i}ve Bayes \\ \textbf{Unsupervised:} \\ Cluster-Based Outlier Detection \\ Attrib-Deviation  }           \\ \hline \rowcolor{babyblueeyes}
Khan et al.~\cite{10}  & $\varoast$ & $\Box$            & \textcolor{green}{\ding{52}} & \textcolor{green}{\ding{52}}  &  \specialcell{ \textit{n}-grams of SQL \\ query Abstraction}  &    \specialcell{n-gram profile comparison \\ mismatches labelled \\ as anomalies} \\ \hline    \rowcolor{babyblueeyes}
Alizadeh et al.~\cite{86} & $\varowedge$                   & $\Box$                  & \textcolor{green}{\ding{52}} & \textcolor{red}{\ding{54}}&  \specialcell{Histograms}  &     \specialcell{Distance b/w histograms}         \\ \hline \rowcolor{babyblueeyes}
Sallam et al.~\cite{18}   & $\varoast$  $\varodot$           & \ding{110}            & \textcolor{green}{\ding{52}} & \textcolor{red}{\ding{54}}   &  \specialcell{Classes / \\ Clusters of \\ Quadruplet}  & \specialcell{ \textbf{Supervised:} \\ Na\"{i}ve Bayes. \\ Multi-labelling classifier \\ \textbf{Unsupervised:} \\ COBWEB}        \\ \hline \rowcolor{babyblueeyes}
Gafny et al.~\cite{102}   & $\varodot$  $\varowedge$         & \ding{110}            & \textcolor{green}{\ding{52}} & \textcolor{red}{\ding{54}}   &  \specialcell{Association Rules}  &     \specialcell{FIM, rule matching}         \\ \hline \rowcolor{babyblueeyes}
Kamra et al.~\cite{250} & $\varoast$          & \ding{110}                  & \textcolor{green}{\ding{52}} & \textcolor{red}{\ding{54}} & quiplets,    &     \specialcell{Na\"{i}ve Bayes}         \\ \hline \rowcolor{babyblueeyes}
Wu et al.~\cite{48}       & $\varoast$ $\varowedge$          & \ding{110}            & \textcolor{green}{\ding{52}} & \textcolor{red}{\ding{54}}     &   \specialcell{Probabilities of \\ Feature's \\ Observed Values }  &   \specialcell{Na\"{i}ve Bayes}       \\ \hline \rowcolor{babyblueeyes}
Santos et al.~\cite{127}  & $\varoast$ $\varodot$ $\varowedge$ & \ding{110}            & \textcolor{green}{\ding{52}} & \textcolor{red}{\ding{54}}  &  \specialcell{ Probability Distribution}  &    \specialcell{Distribution Matching \\ (statistical hypothesis tests \\ i.e. Chi-square, \\ Shapiro-Wilk, \\ Kolmogorov-Smirnov tests)} \\ \hline    \rowcolor{babyblueeyes}
Khan et al.~\cite{288}  & $\varoast$ & $\Box$            & \textcolor{green}{\ding{52}} & Sets of queries &  \specialcell{ query-sets of \\ SQL query Abstraction}  &    \specialcell{Mined Rare query-sets \\ Mismatched frequent query-sets} \\ \hline    \rowcolor{babyblueeyes}
\end{tabular}
\caption{An overview of the characteristics discussed and well-known approaches proposed in the literature. $\varoast$, $\varodot$, and $\varowedge$ represents syntax, data(result), and context-centric features respectively.}
 \label{T2}
\end{table*}
\end{landscape}
    \clearpage
   }

\section{Conclusions} \label{213}
Database intrusion detection, specifically behavioural-based database intrusion detection approaches have seen to be effective in detecting insider attacks.
Two significant aspects in the design of behavioural-based approaches are what level of features ( and in the case of SQL statement then what level of abstraction) are selected for modelling behaviours and the technique (algorithm) selected for constructing profiles. 
The literature contains approaches using syntax-centric, data-centric, or context-centric features or a combination of these features.  To construct profiles, machine learning approach like classification, clustering, as well as distance functions, rule matching algorithms are adopted. 
In the existing literature, pure data-centric, and context-centric approaches are not prevalent.
Some of the approaches are tailored for specific applications, i.e., data warehouses; second, these approaches do not allow for regularly updating normative profiles. 
Majority of these approaches reported in the literature are focused on the detection of a single malicious SQL query in different settings by considering only single SQL query in modelling behaviours~\cite{12,19,86,18,102,127,48}. 
Little attention has been paid on the detection of malicious queries sequences. 
As a single query might not be malicious, but a sequence of SQL queries might be an indication of malicious activity. 
Therefore, such approaches that can detect the malicious sequence of SQL queries are desirable.

\bibliographystyle{plain}
\bibliography{DID_survey} 

\end{document}